\newcommand{\mus}{~\rm \upmu s}
\newcommand{\mum}{~\rm \upmu m}
\newcommand{\neqcm}{~\rm n_{eq}/cm^2}

\newcommand{\e}{~e~\rm ENC}

\documentclass[final,5p,times,twocolumn]{elsarticle} 

\usepackage{amssymb}
\usepackage{upgreek}
\usepackage{graphicx}
\usepackage{lineno}

\journal{Nuclear Instruments and Methods A}

\begin{document}

\begin{frontmatter}



\title{Observations on MIMOSIS-0, the first dedicated CPS prototype for the CBM MVD}

 \author[label1]{M. Deveaux}
 \author[label1]{B. Arnoldi-Meadows}
 \author[label2]{G. Bertolone}
 \author[label2]{G. Claus}
 \author[label2]{A. Dorokhov}
 \author[label2]{M. Goffe}
 \author[label2]{A. Himmi}
 \author[label2]{K. Jaaskelainen}
 \author[label1]{P. Klaus}
 \author[label1]{M. Koziel} 
 \author[label1]{F. Marx}
 \author[label2]{F. Morel}
 \author[label1]{C. M\"untz}
 \author[label2]{H. Pham}
 \author[label2]{M. Specht}
 \author[label2]{I. Valin}
 \author[label1] {J. Stroth}
 \author[label2]{M. Winter}

 \address[label1]{IKF, Goethe University Frankfurt, Max-von-Laue-Str. 1, 60438 Frankfurt/M, Germany }
 \address[label2]{PICSEL Group, IPHC Strasbourg, Rue de Loess 23, 67100 Strasbourg, France}



\begin{abstract}
The Micro Vertex Detector (MVD) of the future Compressed Baryonic Matter (CBM) experiment at FAIR will have to provide a spatial precision of $\sim 5 \mum$ in combination with a material budget of $0.3\%-0.5\% ~X_0$ for a full detector station. Simultaneously, it will have to handle the rate and radiation load of operating the fixed target experiment at an average collision rate of $100~\rm kHz$ ($4-10~\rm  AGeV$ Au+Au collisions) or $10 ~\rm MHz$ (up to $28~\rm GeV$ p-A collisions). The harsh requirements call for a dedicated detector technology, which is the next generation CMOS Monolithic Active Pixel Sensor MIMOSIS. We report about the requirements for the sensor, introduce the design approach being followed to cope with it and  show first test results from a first sensor prototype called MIMOSIS-0 .

\end{abstract}


\end{frontmatter}


\section{Introduction}

The Micro Vertex Detector (MVD) \cite{Paper:CBM-MVD1, CBMMvdPrototype} of the Compressed Baryonic Matter (CBM) Experiment will be composed from four planar detector stations located 5-20 cm downstream the target in target vacuum. A highly heat conductive cooling support made either from TPG or CVD diamond drives the heat of the sensors to liquid cooled heat sinks located outside of the acceptance of the fixed target geometry. This support will host $50\mum$ thin CMOS Monolithic Active Pixel Sensors (CPS) at both sides, which are controlled and read out with low-mass, single-layer flex print cables. The sensors are integrated on both sides to reach a close to 100\% fill factor: The passive area (signal processing circuits) of the sensor from one side is complemented with the active area of the sensor located on the other side.  

The MVD will be operated at an average collision rate of $100~\rm kHz$ ($4-10~  A \rm GeV$ Au+Au collisions) or $10 ~\rm MHz$ (up to $28~\rm GeV$ p-A collisions). A safety margin of a factor of three is foreseen to cope with potential beam fluctuations. This turns into a peak particle rate of $70 ~\rm MHz/cm^2$ and a projected radiation damage of $3\times 10^{13} \neqcm$ plus $3~\rm Mrad$ per year of operation. Thanks to a projected dedicated collimator structure, only a modest number of heavy ion impacts from the beam halo outside of the $5.5~\rm mm$ beam hole are expected. Both the rate and the radiation load are highly position-dependent. The radial symmetry of the radiation field is broken by a high number of delta electrons, which are knocked out from the target by the primary beam and hereafter deflected and focused by the $\sim1 \rm T$ field of the CBM dipole magnet. Delta electrons form up to 90\% of the total occupancy and introduce a complex, 2 dimensional radiation field.

\section{The Design Concept of MIMOSIS}
The MVD will be equipped with a CPS named MIMOSIS. The fundamental design of the pixels matrix is inspired by the ALPIDE sensor \cite{Paper:Alpide}, which was developed for the upgrade of the ALICE ITS upgrade. However, the rate capability and the radiation tolerance of MIMOSIS have to be substantially beyond the ones of ALPIDE, which introduced the need for a dedicated CPS design.

\subsection{Design of the Pixel}
\begin{figure}[t]
\begin{minipage}{0.9\columnwidth}
 \includegraphics[width=\columnwidth]{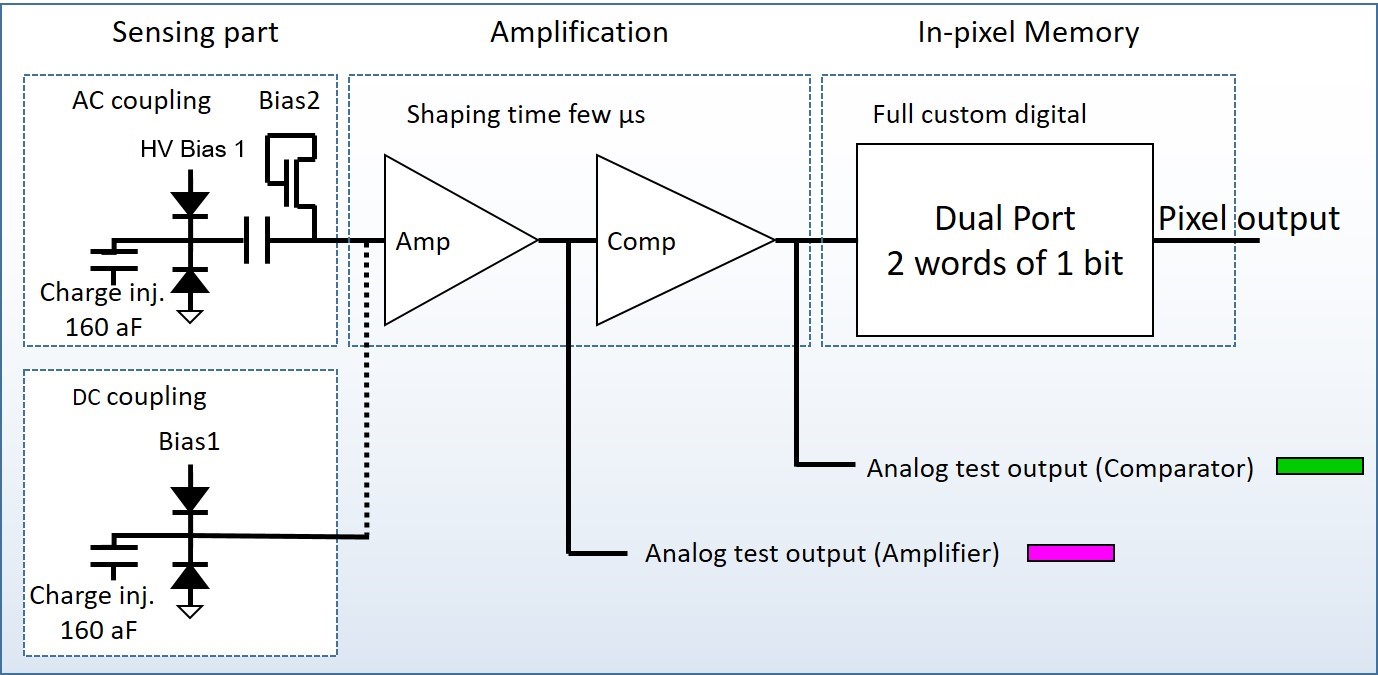}
 \caption{Block diagram of the pixel of MIMOSIS. The analog outputs are shown, the color code (colors online) matches Fig \ref{Fig:AnalogPulse}.}
\label{Fig:Pixel}
\end{minipage}
\begin{minipage}{0.9\columnwidth}
\vspace{3mm}
 \includegraphics[width=\columnwidth]{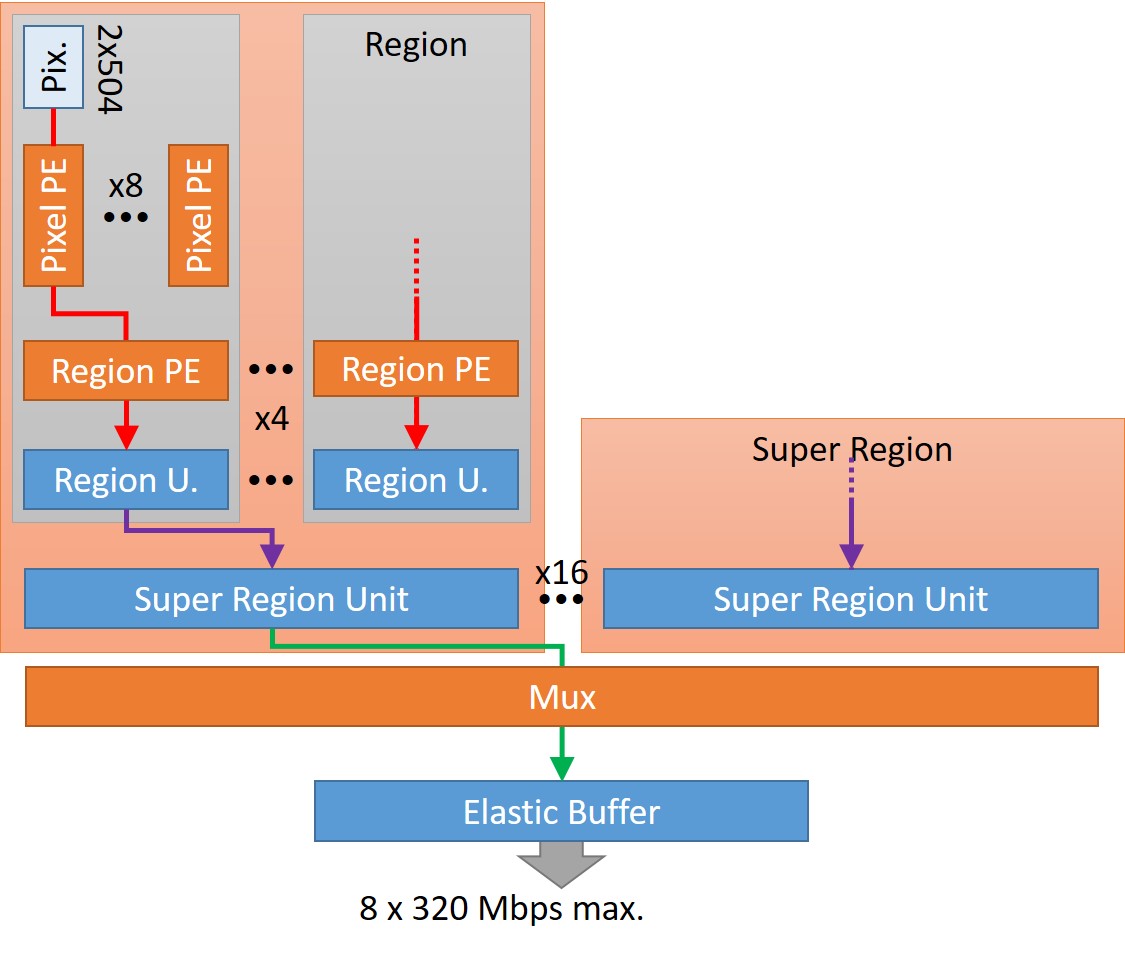}
 \caption{Block diagram of the buffer structure of MIMOSIS.}
\label{Fig:Buffer}
\end{minipage}
\end{figure}

 MIMOSIS will rely on the TOWER/JAZZ 180~nm quad well technology, which allows to integrate a full amplifier-shaper-discriminator chain into each pixel. The pixel matrix will be composed of 1024 columns of 504 pixels with a size of $26.88 \times 30.24 \mum^2$. A block diagram of its layout is shown in Fig. \ref{Fig:Pixel}. The DC coupled version of the pixel is inspired by the related pixel of ALPIDE. However, unlike ALPIDE, which is optimized for being triggered, MIMOSIS will be operated with a continuous readout as required by CBM. Therefore, the trigger logic was removed and replaced by a sample-and-hold logic, which holds a potential signal within the frame time. This tunable frame time is controlled by a global shutter signal and amounts by default to $5\mus$. Once the global shutter arrives, the result of the sample-and-hold circuit is moved to an output buffer and the circuit is cleared. Digital edge detection will be used to avoid a double counting in the frequent case that the time-over-threshold of the  pixel exceeds the frame time. This is required as, unlike to earlier MIMOSA sensors, the dead time of the pixels cannot be neglected. 

Besides the DC version, it is considered to equip MIMOSIS with the AC coupled pixel version also shown in Fig. \ref{Fig:Pixel}. Both pixels are similar but the transistors of the on-pixel amplifier of the AC-pixel is protected from the depletion voltage by means of a capacitor. Therefore, the depletion voltage of the collection diode is not limited by the transistor gates and may be increased to up to $+40 \rm ~V$ while the active medium remains at mass potential. The AC pixel is compatible with the so-called modified \mbox{180~nm} TOWER/JAZZ processes, which are optimized for \mbox{depletion \cite{Paper:ModifiedProcess}}. Therefore, it will be most likely possible to reach a full depletion of the sensor, which is expected to increase the radiation tolerance of the sensor substantially.

\subsection{Buffer structure}

The buffer structure of MIMOSIS is displayed in Fig. \ref{Fig:Buffer}. It has to satisfy the need to read out any recorded hit without reducing the data load by means of a trigger system.
The readout of the pixels is done with a priority encoders, which each serve two columns and operate at a readout frequency of $20~\rm MHz$. The time resolution of the system is determined by the frame length. The latter is foreseen to amount to $5 \mus$. Other values may be chosen within some limits on expense of a lower hit rate capability\footnote{Choosing a faster time resolution will load the data busses with additional Poisson fluctuations of the particle/data flow and additional network overhead. Reducing the time resolution will challenge the depth of the data buffers.}. The maximum number of hits recorded from a double column corresponds to the number of clock cycles within a frame time. If the number of hits exceeds this number, the non-read hits are discarded.

The data of eight double columns will be provided to a so-called region buffer via a data concentration unit. This unit identifies groups of up to four neighbouring fired pixels in order of readout. This corresponds to a 3D cluster finding within the double column but it is not suited to identify clusters spanning more than one double-column. The region buffer is designed as a double buffer with one element reading the priority encoded data from the double columns with a $16~\rm bit \times 20~\rm MHz$ bus while the second buffer is writing the data of the previous frame to the next stage. The region buffer adds a dedicated ``region'' header, which reduces the number of bits required for encoding the column number of the hits. The depth of the region buffer is 100 words, which corresponds to the theoretical maximum output of a single double column but is typically shared among those columns. This sharing represents a first averaging of occupancy fluctuations as caused by Poisson fluctuations. 

The 16 super regions concentrate the data from four regions each via a priority encoded $32 ~\rm bit \times 40~MHz$ bus. Its depth is 256 words and its purpose is load balancing and preparing the data for a transport via the main $256 \rm bit \times 40~MHz$ bus. This bus will send the data via an idle word removing unit (not shown in Fig. \ref{Fig:Buffer}) toward one central elastic buffer. The idle word removal aims to eliminate potential idle words, which were created to fill up the messages in the buses with a width beyond \mbox{16 bit}. It is suited to process a maximum of 3200 words per $5\mus$. Hereafter, the data is written into an elastic output buffer with an input bandwidth of 3200 words per $5\mus$ and a depth of 16384 words, which is sufficient to store the peak data rate corresponding to 5 frames. The elastic buffer forms the central element of the temporal and spatial load balancing. It sends its data to up to eight differential $320~\rm Mbps$ output buffers. In case of a low data load, the number of output buffers may be reduced in order to save power and the material budget of traces.

\section{First Results from Prototyping}

\begin{figure}[t]
\begin{minipage}{0.9\columnwidth}
 \includegraphics[width=\columnwidth]{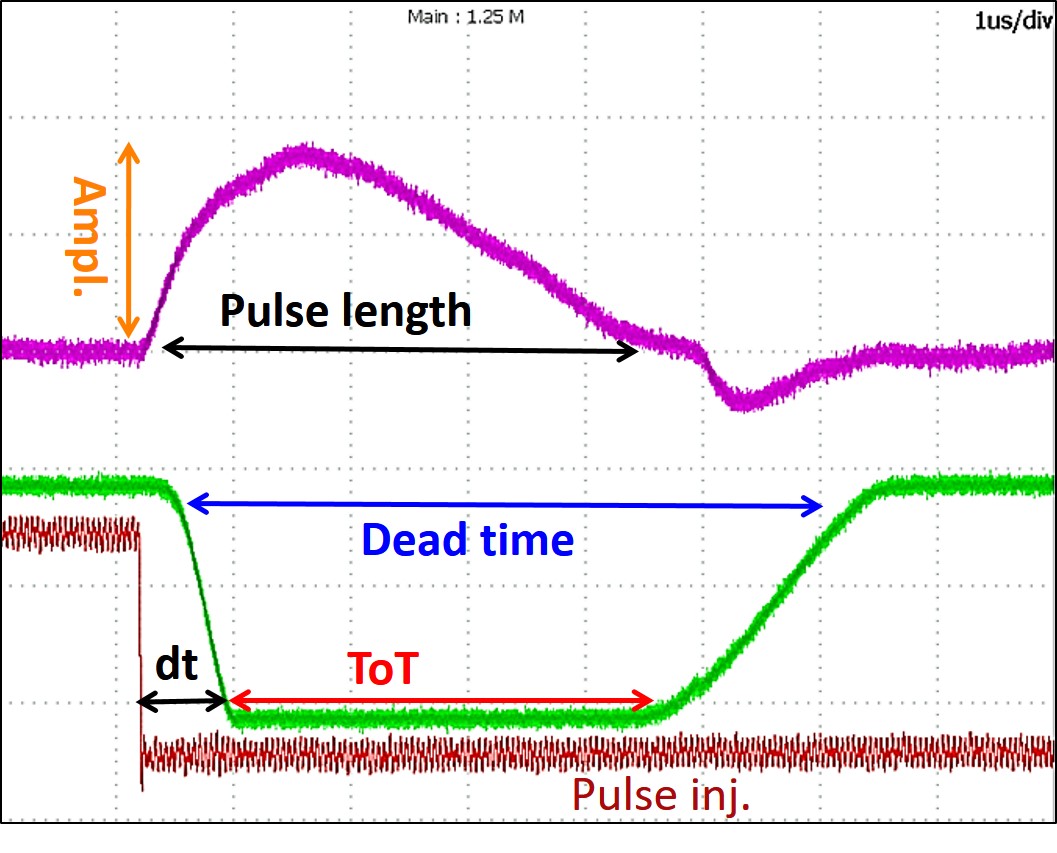}
 \caption{Oscilloscope screen shot of the analog pulse shape of a DC pixel of MIMOSIS-0. The upper line represents the amplifier output, the lower line the comparator output. Moreover the pulse injection is shown. The voltage units are arbitrary due
to intermediate buffering.}
\label{Fig:AnalogPulse}
\end{minipage}
\begin{minipage}{0.9\columnwidth}
 \includegraphics[width=\columnwidth]{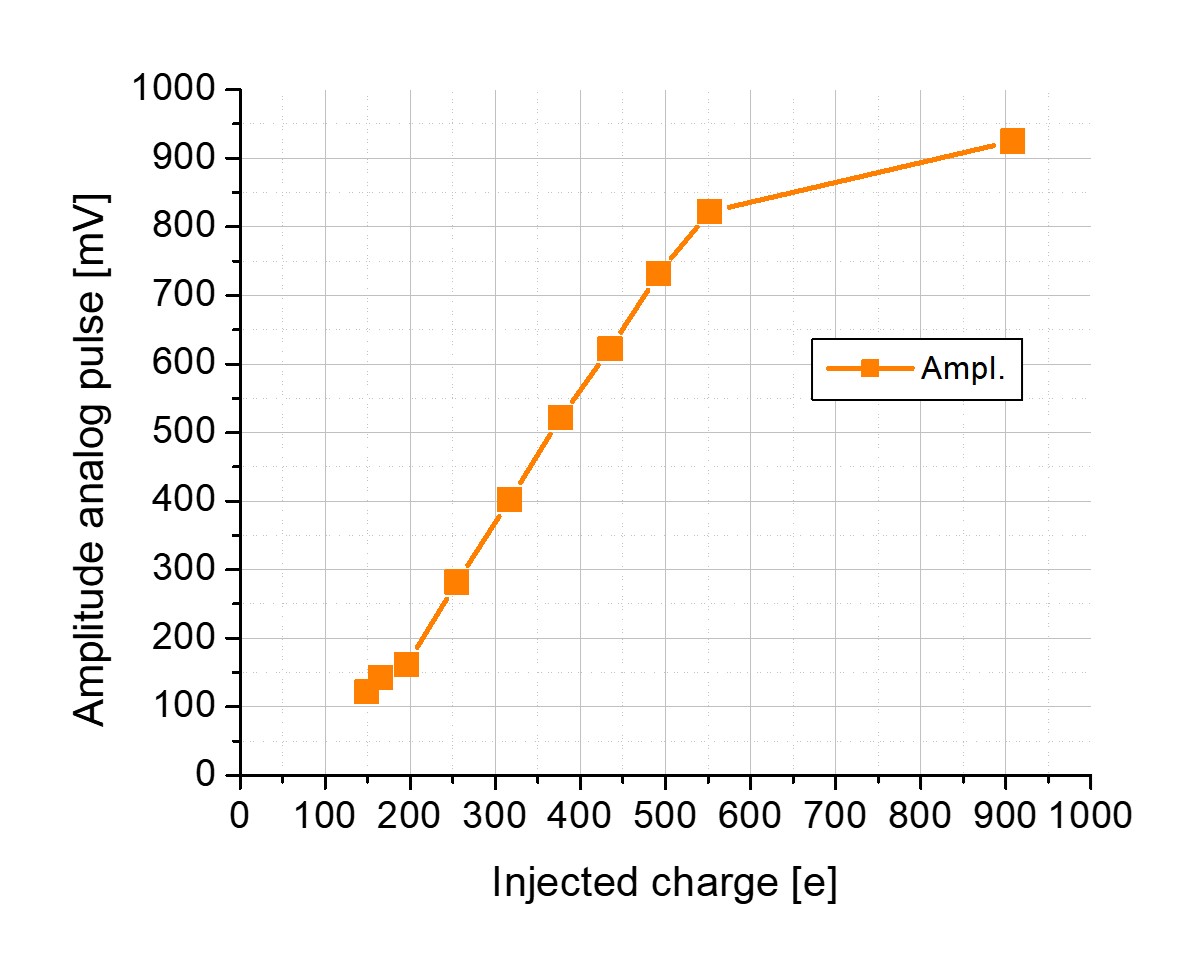}
 \caption{Pulse height of the amplifier signal as function of the injected charge.}
\label{Fig:Gain}
\end{minipage}
\end{figure}

The final MIMOSIS sensor will be build based on a dedicated R\&D program, which is currently considered to consist of three prototype submissions. The first prototype, MIMOSIS-0, was submitted in 2018 and is being tested. The sensor consists of 16 double columns with $2\times 504$ pixels each. Half of the pixels were realized with AC and DC coupling, respectively. The priority encoder used for the pixel readout as well as the DACs required for steering the pixels were implemented. A selected number of pixels were realized with a dedicated analogue readout line, which allows to observe the analogue signals within the in-pixel amplification chain directly. As displayed in the simplified block diagram shown in Fig.~\ref{Fig:Pixel}, those outputs allow to monitor the output of the amplifier and the discriminator, respectively. Probing the input of the amplifier was also tried but remained unsuccessful. 

\subsection{Properties of the Amplification Chain}
Fig. \ref {Fig:AnalogPulse} shows the signal shape observed for an pulse injection of roughly $250~e$. The upper line corresponds to the output of the amplifier and the lower line to the output of the discriminator. The output was used to measure the gain of the amplifier by modifying the injected charge in steps of roughly $6~e$ per $\rm ADU$ and measuring the amplitude of the amplifier signal. Note that by doing so, we assume the capacity of the charge injection system to amount precisely its nominal value of $C=160~\rm aF$. As seen in Fig. \ref{Fig:Gain}, the amplifier is found linear in a range between roughly $150~e$ and $600~e$. Below this value, the gain is reduced, which complicates choosing substantially lower thresholds independently of the noise. Above $600~e$, the amplifier is intentionally saturated in order restrict the pulse duration. 

\begin{figure}[t]
\begin{minipage}{0.9\columnwidth}
 \includegraphics[width=\columnwidth]{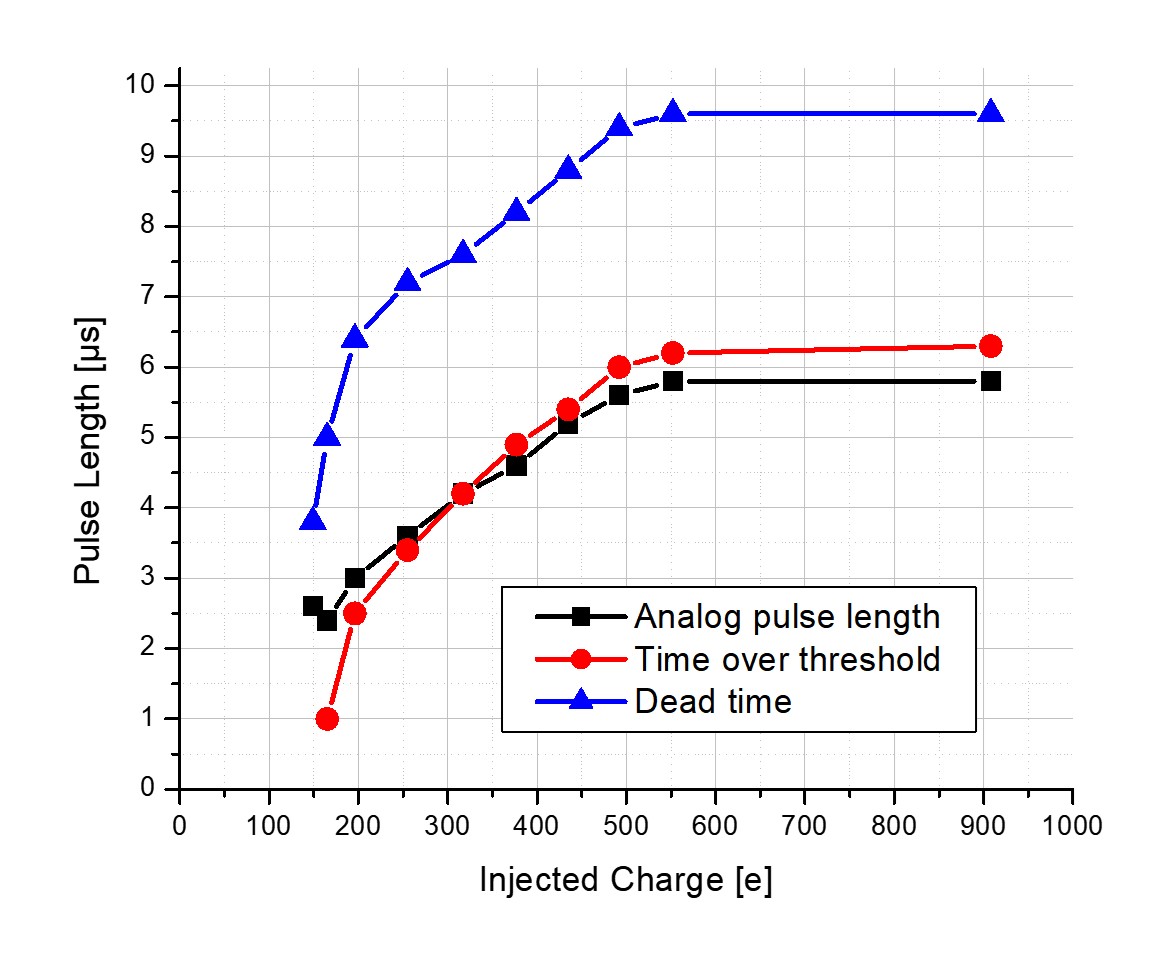}
 \caption{Pulse length of a DC pixel as a function of the injected charge. See Fig. \ref{Fig:AnalogPulse} for the definition of the quantities.} 
\label{Fig:PulseLength}
\end{minipage}
\begin{minipage}{0.9\columnwidth}
 \includegraphics[width=\columnwidth]{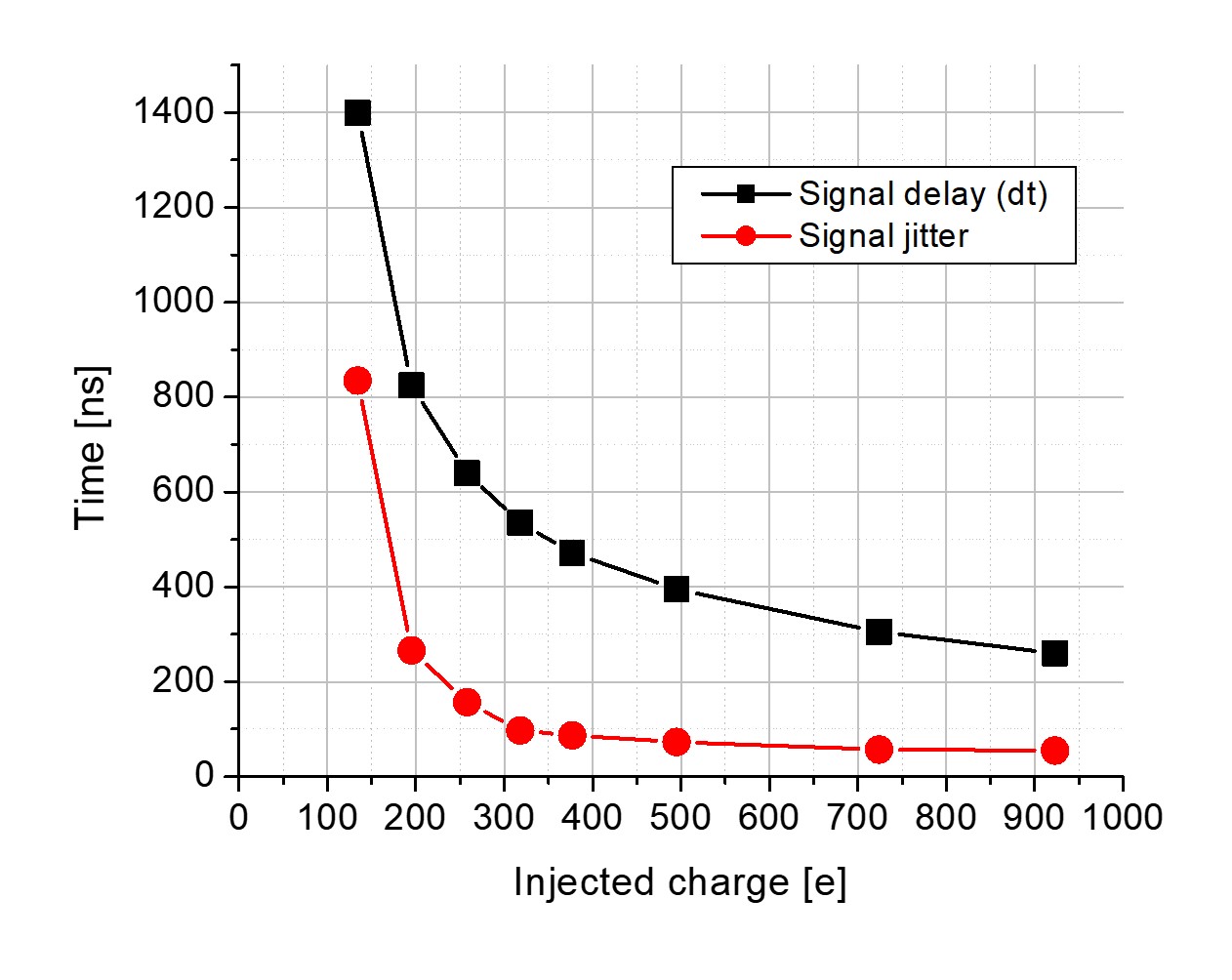}
 \caption{Delay and jitter (at 90\% of the maximum amplitude) of the comparator signal with respect to the signal injection as a function of the injected charge.}
\label{Fig:TimeWalk}
\end{minipage}
\end{figure}

The pulse length was also measured as a function of charge. As displayed in Fig.~\ref{Fig:PulseLength}, the pulse length increases with the charge and may reach $\sim 6 \mus$ for the analogue pulse and the time-over-threshold while the dead time may reach $\sim 10 \mus$. The precise values vary by few $100 ~\rm ns$ depending on the individual pixel and depend moreover to some limited extent on the detailed settings of the eight current and voltage sources, which steer the pixels. It is worth mentioning that, due to space constraints, all pixels share common voltage sources. Therefore, an individual tuning of the pixels is not feasible. 

The delay of the discriminator response was measured with respect to the moment of the pulse injection (see Fig. \ref{Fig:TimeWalk}). One observes a time walk (variation of the delay as function of the signal amplitude) of $\sim 1.2\mus$ and a jitter of $0.85\mus$, which determines the theoretical optimal time resolution of the analogue chain. Both parameters shrink rapidly with increasing amplitude. For hits providing a still modest signal of $>300~\e$, the time resolution might already be as good as $0.6 \mus$, which could be conceptually exploited for limited occupancies by reducing the frame time. In any case, one should keep in mind that CPS form pixel clusters and that the neighbouring pixels collect less signal than the seed pixel. In consequence, their signal indication might be delayed with respect to the one of the seed pixel, which must be considered in data analysis.

\subsection{Pixel Noise}

The noise of the pixels was measured at room temperature by means of an S-curve scan. Doing this scan was complicated by the non-linearity of the amplifier, which reaches its amplification regime for signals above $\sim 100~\e$ only. 
The noise was therefore measured by setting a fixed average detector threshold and by varying the injected charge of the pixels in a range nearby this threshold. This procedure was carried out on the DC and AC pixels
of the CE18 test structure, which hosts the same pixel amplifier structures in the absence of a digital logic. Moreover, the DC pixel of MIMOSIS-0 was tested. The AC pixel of MIMOSIS show a coupling with the digital signals of the priority encoder.
This feature is understood and will be fixed in the next design iterations but hampered us from using the digital output as required for  S - curve scans.
Note that the noise values provided include the noise of the charge injection system.

\begin{table}
	\centering
		\begin{tabular} {|c|c|r|r|r|r|}
		\hline
		  Sensor & Pixel &BB & Thresh. & FPN \\ 
			\hline
			MIMOSIS-0 & DC& 0V &$155 ~ e$  & $15 ~ e$ \\
			MIMOSIS-0 & DC& 0V &$120 ~ e$  & $20 ~ e$ \\
			MIMOSIS-0 & DC& -1V&$155~e$ & $7 ~e$ \\
			\hline
			CE18 & DC& -1V&$150~e$ & $9 ~e$   \\
			CE18 & AC& -1V&$150~e$ & $7 ~e$   \\
			\hline

		\end{tabular}
	\caption{Selected results on the noise of MIMOSIS-0 and the CE18 test structure. All noise values are given in units of $\e$
	and hold for room temperature. }
	\label{tab:NoiseResults}
\end{table}

The noise is found to be dominated by the Fixed Pattern Noise (FPN), which denotes the offset of the pixels w.r.t the common 
threshold. Selected results on the FPN are displayed in Tab. \ref{tab:NoiseResults}. One observes that the fixed pattern noise of the pixels remains
in a satisfactory range and it is further reduced if a mild back bias of $-1~\rm V$ is applied to the p-well structures of the sensor.
Increasing this voltage further shows no substantial effect, which might be related to a missing experimental sensitivity. The thermal
noise of the pixels was measured with CE18 as it is in the order of the discretization steps of the DACs used for signal injection
in MIMOSIS-0. We observe a thermal noise of $\lesssim 9 e ~\rm ENC$ for the DC pixel and of $ \lesssim 3 e ~\rm ENC$ for the AC pixel.

\subsection{Response to $^{55}$Fe}

\begin{figure}[t]
\begin{center}
 \includegraphics[width=0.9\columnwidth]{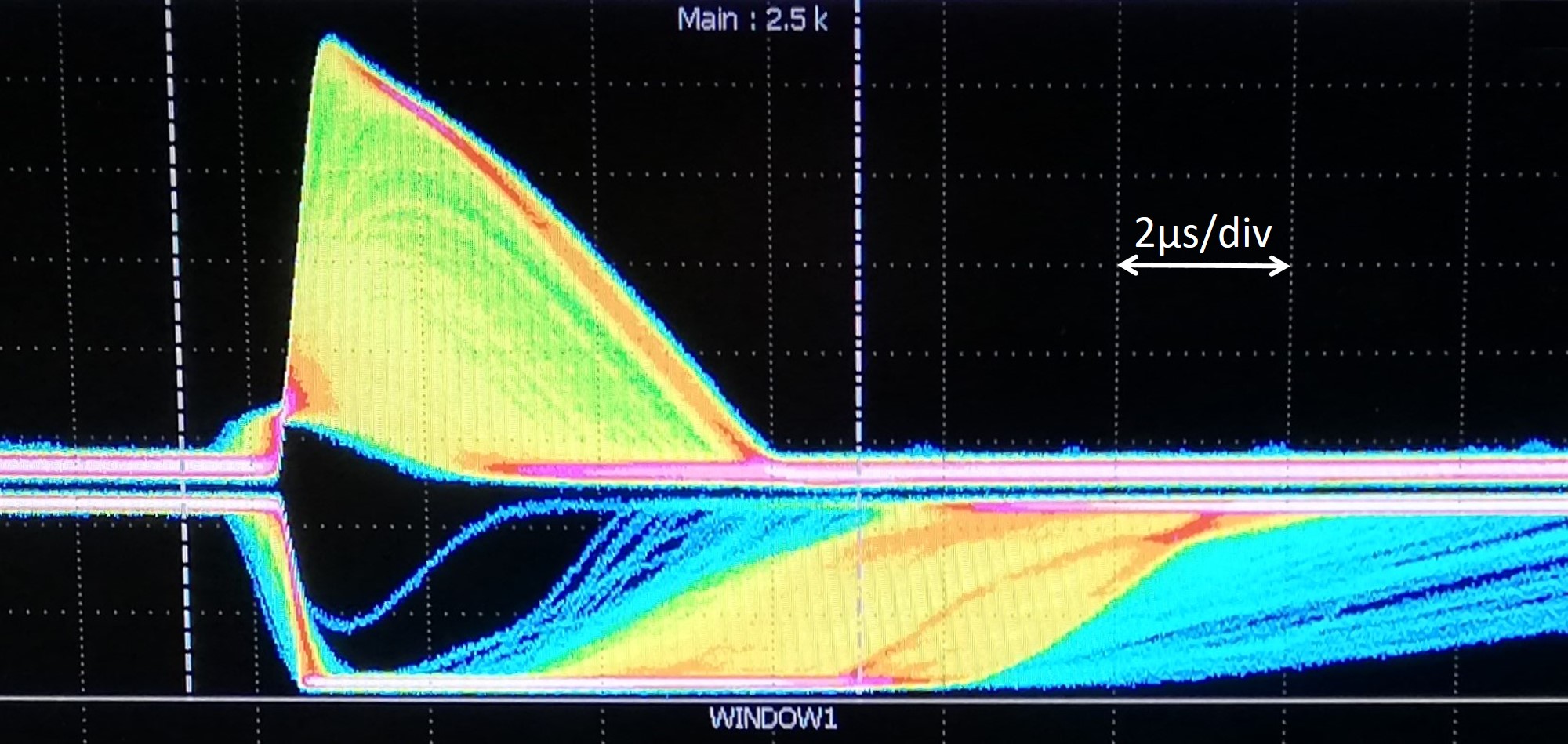}
\end{center}
 \caption{Response of a DC-pixel of MIMOSIS-0 to $5.9~\rm keV$ photons from a $^{55}$Fe-source as seen in an accumulated plot with an oscilloscope (blue= low population, red/white= high population). The signals are displayed
in analogy to Fig. \ref{Fig:AnalogPulse}.} 
\label{Fig:Fe55}
\end{figure}

The sensitivity of MIMOSIS-0 to radiation was tested by illuminating the DC pixels with the $5.9~\rm keV$ photons from a $^{55}$Fe-source. The signal of one pixel with analogue output was accumulated over
numerous hits by means of an oscilloscope. A related screenshot is displayed in Fig. \ref{Fig:Fe55}. One observes that in many
cases, the $1640~e$ generated by the photon exceed the saturation limit of the pixel. However, a significant number of 
entries with low signal charge are visible, which may be related to hits occurring far from the collection diode (e.g. in the neighbouring pixel). 
Due to both features, little further quantitative information may be extracted from the the experiment.  The distribution of the 
hits over the pixel matrix was also tested and remained without unexpected features. 

\section{Summary and conclusion}

The CBM Micro Vertex Detector will be equipped with a dedicated CMOS Monolithic Active Pixel Sensor named MIMOSIS. The sensor will have to fulfil requirements of $5\mum$ spatial resolution, $50\mum$ thickness in combination with a rate of up to $70~\rm MHz/cm^2$ and a time resolution of $\sim 5 \mus$. To fulfill this task, MIMOSIS will be equipped with pixels hosting a full amplifier-shaper-discriminator chain and a priority encoded readout, which was inspired by the one of the ALPIDE sensor used for the ALICE ITS upgrade. Modifications of MIMOSIS with respect to ALPIDE consist in replacing the trigger oriented readout of ALPIDE by a high performance continuous readout relying on a massively parallel data concentration unit, which is realized with a three layer buffer structure. Moreover, optional AC coupled pixels will allow for reaching a full depletion of the active volume.

Preliminary tests of a prototype called MIMOSIS-0 demonstrated a successful integration of the analogue electronics of pixels with AC and DC coupled preamplifiers, the priority encoder and the slow control units. Both pixels show a low noise allowing for a threshold setting of $<150~e$ and the noise may be further reduced by applying a mild, $-1~\rm V$ back bias to the p-well structures. The pulse shapes and lengths of the in-pixel amplification chain was studied and it is concluded that it is conceptually suited to reach a $\sim 1\mus$ time resolution in combination with a dead time of $\sim 10 \mus$, which is faster than the ambitioned frame time of $5 \mus$. MIMOSIS-0 is currently being tested for its radiation tolerance and first results are expected in a close future.
Moreover, a first reticle size prototype is being prepared for a submission in 2019.

\section*{Acknowledgements}
This work was supported by BMBF(05P15RFFC1), GSI and HIC for FAIR.





\end{document}